\documentclass[onecolumn,showpacs,nobibnotes,nofootinbib]{revtex4}

\usepackage{amsmath,amssymb}
\usepackage{graphicx}
\usepackage{subfigure}

\newcommand{\x}{\mathbf{x}}
\newcommand{\y}{\mathbf{y}}
\newcommand{\z}{\mathbf{z}}

\newcommand{\abs}[1]{\left|#1\right|}
\newcommand{\ab}{\bar\alpha}
\newcommand{\as}{\alpha_s}
\newcommand{\avg}[1]{\left\langle #1 \right\rangle}
\newcommand{\erfc}{\text{erfc}}

\begin{document}

\title{Fluctuations effects in high-energy evolution of QCD}
\author{G. Soyez\footnote{on leave from the fundamental theoretical physics 
group of the University of Li\`ege.}}
\email{gsoyez@spht.saclay.cea.fr}
\affiliation{SPhT \footnote{URA 2306, unit\'e de recherche associ\'ee au CNRS.}, 
CEA Saclay, B\^{a}t. 774, Orme des Merisiers, F-91191 Gif-Sur-Yvette, France}

\begin{abstract}
Recently, Iancu and Triantafyllopoulos have proposed a hierarchy of evolution equations in QCD at high energy which generalises previous approaches by including the effects on gluon number fluctuations and thus the pomeron loops. In this paper, we present the first numerical simulations of the Langevin equation which reproduces that hierarchy. This equation is formally the Balitsky-Kovchegov equation supplemented with a noise term accounting for the relevant fluctuations. In agreement with theoretical predictions, we find that the effects of the fluctuations is to reduce the saturation exponent and to induce geometric scaling violations at high energy.
\end{abstract}

\maketitle

\section{Introduction}\label{sec:intro}

The question of the high-energy limit of QCD is a longstanding problem. The resummation of the leading logarithmic contributions at high-energy, well-known as the BFKL equation \cite{bfkl}, predicts a power growth $s^\omega$ of the scattering amplitude $T$. In the large-$N_c$ approximation, the resummation of these leading logarithmic contributions is equivalent \cite{mueller} to the evolution of a wavefunction considered as a system of $q\bar q$ dipoles.

At sufficiently high energy, this violates the Froissart bound and the unitarisation effects \cite{glr} have to be taken into account. In the large-$N_c$ approximation, it has been shown that the relevant way to introduce these unitarisation effects is to include multiple-scattering effects in the collision between external dipoles and evolved target. This leads to an infinite hierarchy of equations \cite{balitsky,jimwlk} taking into account linear BFKL growth and saturation effects in the high-density regime. In the mean-field approximation, valid {\em e.g.} for dense targets such as heavy nuclei, the infinite hierarchy resumes to a single non-linear equation known as the Balitsky-Kovchegov (BK) equation \cite{balitsky,kovchegov}.

Recently, one has noticed \cite{ms,imm} that, for dilute systems, one also have to include the effects of fluctuations of the number of gluons. It has been subsequently recognised, by Iancu and Triantafyllopoulos \cite{it}, that these contributions are not included in the original Balitsky/JIMWLK hierarchy but can be added to it. The additional terms describe gluon splittings in the target wavefunction and act as a seed for the pomeron loops. The resulting infinite set of equations is equivalent, up to a coarse-graining approximation, to a stochastic Langevin equation which formally have the form of the BK equation supplemented with a noise term responsible for fluctuations \cite{it}.

In the saddle-point approximation, Munier and Peschanski have shown \cite{Munier1} that the BK equation is equivalent to the Fisher-Kolmogorov-Petrovsky-Piscounov (FKPP) equation \cite{KPP} well-known in statistical physics. As a consequence of this equivalence, one knows that the asymptotic solutions of the BK equation takes the form of a traveling wave and leads to geometric scaling \cite{geometric,Munier1} {\em i.e.} to the fact that the scattering amplitude depends only on $k^2/Q_s^2(Y)$ where $k$ is the transverse momentum of the dipole and $Q_s(Y)$ is the rapidity-dependent saturation scale. If we includes the noise term, the equation becomes equivalent to the stochastic FKPP (sFKPP) equation. In this type of equation, geometric scaling is observed event by event but the averaged amplitude leads to violations of geometric scaling as the events are diffused around their average position.

The equivalence between QCD and statistical-physics equations is actually an approximation. If we go beyond the saddle point approximation, {\em i.e.} if we look at the exact QCD evolution equation, we get additional non-local contributions. These can be studied using the same techniques, leading to slightly modified asymptotic properties and to a different infrared behaviour. In addition, the analytic results \cite{imm,it} obtained in statistical physics are only valid for asymptotically small values of the strong coupling constant $\alpha_s$. Only a numerical analysis can give relevant results in the physical domain. So far the only numerical studies of the QCD equations including fluctuations \cite{anna} has been restricted to models without transverse coordinate dependence.

The aim of this paper is to study numerically the properties of the solutions of the Langevin equation describing the QCD amplitude and to compare with the results obtained from BK evolution. We shall check, in the case of the QCD Langevin equation and for real-life values of the physical parameters, that the effects of the fluctuations are well predicted by the sFKPP-inspired approximation \cite{it} at large rapidities and in the tail of the front. Basically, we are going to study the effect of fluctuations on the dispersion of the events, the $Y$ dependence of the saturation scale and the form of the wavefront. Let us emphasise that our studies are directly related to the QCD equation, hence we do not solve the sFKPP equation or any other model such as the effect of a cut-off \cite{brunet} on the amplitude.

The structure of the paper is as follow. We shall first explain the basic considerations leading to the hierarchy of evolution equations including fluctuations and introduce the Langevin equation. We shall then then expose the results on the asymptotic solutions obtained from the equivalence with statistical physics equations. We then explain how we solve numerically the QCD equation and the results obtained from this analysis. We finally conclude and give the perspectives of this paper.

\section{Fluctuations in QCD evolution}\label{sec:bdep}

\begin{figure}
\subfigure[linear term]{\includegraphics{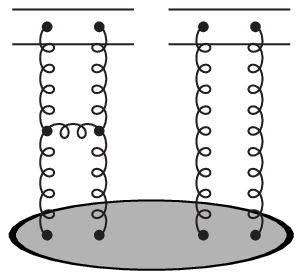}}
\subfigure[fan diagram]{\includegraphics{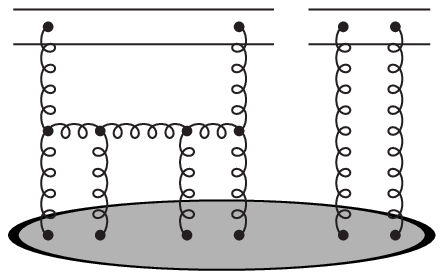}}
\subfigure[fluctuation]{\includegraphics{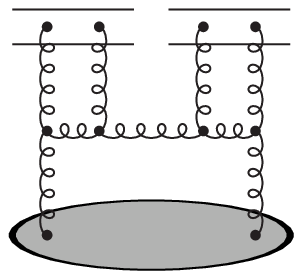}}
\caption{These figures show the contribution to the evolution of $T^{(2)}$: a linear growth proportional to $T^{(2)}$, a merging term proportional to $T^{(3)}$ and a splitting term proportional to $T$.}
\label{fig:diagr}
\end{figure}

In this section, we briefly recall the basis steps yielding to the Langevin equation for the rapidity evolution of QCD amplitudes. The limit we want to consider is large $Y=\log(s)$, and $\alpha_s$ small enough to allow for a perturbative treatment. We shall work in the framework of the large-$N_c$ approximation and introduce $\ab = \frac{N_c \alpha_s}{\pi}$. In order to study the rapidity evolution of a target, we shall probe it with a set of projectile made of $q\bar q$ dipoles. 

It is now well-known \cite{glr,glr+} that, due to nonlinear unitarisation effects, it is not sufficient to consider interaction with one single dipole under the form of a BFKL ladder. In other words, it is not sufficient to consider a simple two-gluon exchange in the $t$ channel and the associated amplitude $T$, but one has to take into account the higher-order correlations $\langle T(\x_1,Y)\dots T(\x_n,Y)\rangle$. These nonlinear effects introduces \cite{balitsky,jimwlk} in the evolution of the $n$ dipole system a contribution from the $n+1$ dipole sector. This translates the fact that unitarity effects in dense systems take the form of recombinations.

It has recently been shown \cite{ms,imm,it} that in addition to the unitarity corrections relevant in the dense regime, it is also of prime importance to take into account gluon number fluctuation terms if one wants an acceptable description of the dilute systems. If one put all these effects together, Iancu and Triantafyllopoulos obtain \cite{it} an infinite hierarchy of equations for all dipole correlations. If $T^{(k)}(\x_1,\y_1;\dots;\x_k,\y_k;Y)$ is the correlation for the interaction with $k$ dipoles of coordinates $(\x_1,\y_1)$, ..., $(\x_k,\y_k)$, this hierarchy takes the form
\begin{eqnarray}\label{eq:hier}
\lefteqn{\partial_Y T^{(k)}(\x_1,\y_1;\dots;\x_k,\y_k;Y)}\nonumber\\
 & = & \ab \int d^2z\, \sum_{j=1}^k\frac{(\x_j-\y_j)^2}{(\x_j-\z)^2(\z-\y_j)^2}
\left\lbrack
T^{(k)}(\x_1,\y_1;\dots;\x_j,\z;\dots;\x_k,\y_k;Y)
+T^{(k)}(\x_1,\y_1;\dots;\z,\y_j;\dots;\x_k,\y_k;Y)
\right. \nonumber\\
 &   & \phantom{\ab intz d^2z\, sum(\x_j-\z)^2(\z-\y_j)^2}
\left.
-T^{(k)}(\x_1,\y_1;\dots;\x_k,\y_k;Y)
-T^{(k+1)}(\x_1,\y_1;\dots;\x_j,\z;\z,\y_j;\dots;\x_k,\y_k;Y)
\right\rbrack \nonumber\\
 & + & \ab\alpha_s^2\kappa\, \sum_{j=1}^{k-1}\frac{(\x_j-\y_j)^2(\x_{j+1}-\y_{j+1})^2}{(\x_j-\y_{j+1})^2}
\,T^{(k-1)}(\x_1,\y_1;\dots;\x_j,\y_{j+1};\dots;\x_k,\y_k;Y)\,\delta^{(2)}(\y_j-\x_{j+1}),
\end{eqnarray}
where $k$ runs over all positive integers. These equations are obtained if one makes the approximation that the amplitude is proportional to $\alpha_s^2(\x\!-\!\y)^4n(\x,\y)$ where $n$ is the gluon number density. The proportionality constants appears as the unknown factor $\kappa$ in the fluctuation term. This approximation is correct in the BFKL or in the dilute regime. The constant $\kappa$ is expected to be of order 1. In fact, this approximation has been relaxed in subsequent publications (see second paper in Ref. \cite{it} and \cite{msw}) but the non-locality of the ensuing equation is too complicated to be accounted for in numerical simulations. This is why throughout this paper we shall restrict ourselves to the simplified hierarchy \eqref{eq:hier}. In fact, in what follows we shall consider a simplified version of equation \eqref{eq:hier} obtained after a coarse-graining in impact-parameter space.

To make things clearer, let us discuss the evolution equation for the correlations of two dipoles $T^{(2)}$
\begin{eqnarray}
\lefteqn{\partial_Y T^{(2)}(\x_1,\y_1;\x_2,\y_2;Y)}\nonumber\\
 & = & \ab \int d^2z\, \frac{(\x_1-\y_1)^2}{(\x_1-\z)^2(\z-\y_1)^2}
\left\lbrack
T^{(2)}(\x_1,\y_1;\x_2,\z;Y)+T^{(2)}(\x_1,\y_1;\z,\y_2;Y)-T^{(2)}(\x_1,\y_1;\x_2,\y_2;Y)
\right. \nonumber\\
 &   & \phantom{\ab intz d^2z\, (\x_j-\z)^2(\z-\y_j)^2}
\left.
-T^{(3)}(\x_1,\y_1;\x_2,\z;\z,\y_2;Y) + (1 \leftrightarrow 2)
\right\rbrack \nonumber\\
 & + & \ab\alpha_s^2\kappa\, \frac{(\x_1-\y_1)^2(\x_2-\y_2)^2}{(\x_1-\y_2)^2}
\,T^{(1)}(\x_1,\y_2;Y)\,\delta^{(2)}(\y_1-\x_2).
\end{eqnarray}
We distinguish three different types of contributions to the rapidity evolution of $T^{(2)}$. The first one is the BFKL linear term, proportional to $T^{(2)}$ itself. It corresponds to the diagram of figure \ref{fig:diagr}(a) and contributes to the elaboration of a BFKL ladder. Then, the term proportional to $T^{(3)}$, shown in figure \ref{fig:diagr}(b), resums fan diagrams, responsible for the saturation effects. Finally, the last term (figure \ref{fig:diagr}(c)) corresponds to the fluctuations and introduces a contribution from $T$ in the evolution of $T^{(2)}$. 

It is interesting to check the regimes in which these terms are expected to be important. This is easily done by considering the power of $\alpha_s$ associated with each graph. Since $T$ represents at least a two-gluon exchange, each power of $T$ carries 4 vertices {\em i.e.} a factor $\alpha_s^2$. One thus easily see that the terms proportional to $T^{(2)}$ or $T^{(3)}$ both carry an additional factor $\alpha_s$ while the fluctuation term, proportional to $T$ has an extra factor $\alpha_s^3$. This means that the merging contribution becomes important when $T^{(k)}\sim T\sim 1$. It takes into account saturation effects in the dense regime. Moreover, the fluctuation term becomes important when $\alpha_s^2 T \sim T^{(2)}$, {\em i.e.} when $T\sim \alpha_s^2$. One thus expects large effects due to the fluctuations in the dilute regime. As an extreme case, if one starts with a single dipole, we have $T^{(k)}=0$ for $k > 1$ and the fluctuation term is needed to obtain a nonzero value for the correlations.

Note that in the hierarchy, the physical roles of the QCD parameters $\ab$ and $\alpha_s^2$ are very different. While $\ab$ always multiplies $Y$ and hence sets the scale for the rapidity evolution, $\alpha_s^2$ measures the strength of the elementary dipole-dipole interaction and thus sets the scale for the fluctuations. In fact, in the numerical simulations, we found convenient to fix the value of the QCD coupling constant (fixing simultaneously the values of $\ab$ and $\alpha_s^2$) and to use the freedom in the choice of the parameter $\kappa$ as a method to vary the strength of the dipole interaction.

If we neglect the fluctuation term in this equation, one recovers the large-$N_c$ version of the JIMWLK hierarchy \cite{jimwlk}, which is coherent with the fact that it is then expected to describe dense systems such a heavy nuclei. It is also interesting to notice that, while the JIMWLK equation at large $N_c$ resums just fan diagrams, the complete hierarchy including the fluctuations also allows for pomeron loops. As a simple example, in two steps of evolution, $T^{(1)}$ can fluctuate into $T^{(2)}$ and merge back into $T^{(1)}$.


If we perform a coarse-graining approximation in impact parameter, the fluctuation term becomes local and we can perform the Fourier transform to momentum space (see Ref. \cite{it} for technical details). By applying the same methods as those exposed in appendix \ref{app:equiv}, the resulting hierarchy of equation in momentum space can be rewritten under the form of a Langevin equation
\begin{equation}\label{eq:langevin}
\partial_Y T(k,Y) = \ab \int_0^\infty \frac{dp^2}{p^2}\,
\left[ 
\frac{p^2 T(p,Y) - k^2T(k,Y)}{\abs{p^2-k^2}} + \frac{k^2 T(k,Y)}{\sqrt{4p^4+k^4}}
\right] -\ab T^2(k) 
+ \ab \sqrt{2\kappa \as^2 T(k)}\, \nu(k,Y),
\end{equation}
where the noise $\nu(k,Y)$ satisfies
\[
\avg{\nu(k,Y)} = 0 \qquad \text{and} \quad 
\avg{\nu(k,Y)\nu(k',Y')} = \frac{1}{\ab\pi}\delta(Y-Y')\,k\,\delta(k-k').
\]
This equation has to be understood as the continuum limit with the Ito prescription which means that if we perform rapidity steps of size $\varepsilon$, labelled by an index $j$, it can be rewritten
\begin{equation}\label{eq:ito}
\frac{T_{j+1}(k)-T_j(k)}{\varepsilon}
 = \ab K\otimes T_j(k) - \ab T_j^2(k) 
+ \ab \sqrt{2\kappa \as^2 T_j(k)}\, \nu_j(k),
\end{equation}
where $K\otimes T$ denotes the convolution with the linear kernel and the noise $\nu_j$ satisfies
\[
\avg{\nu_j(k)} = 0 \qquad \text{and} \quad 
\avg{\nu_j(k)\nu_{j'}(k')} = \frac{1}{\varepsilon\ab\pi}\delta_{jj'}k\delta(k-k').
\]

\section{Traveling waves and asymptotic properties}\label{sec:trav}

Now that we have presented the equation describing high-energy evolution of QCD amplitude, let us quote what are the expected properties of its solutions. 

The linear term in \eqref{eq:langevin} can be rewritten $\chi(-\partial_L) T$, with $L = \log(k/k_0)$ and $\chi(\gamma) = 2 \psi(1)-\psi(\gamma)-\psi(1-\gamma)$ is the eigenvalues of the BFKL kernel. If one expand this operator to second order around $\gamma=1/2$, the equation \eqref{eq:langevin} becomes after a linear change of variables and a rescaling of $T$
\begin{equation}\label{eq:sfkpp}
  \partial_t u(x,t) = \partial_x^2 u(x,t) + u(x,t) - u^2(x,t) + \sqrt{2\eta u(x,t)} \nu(x,t).
\end{equation}
This equation is the stochastic Fisher-Kolmogorov-Petrovsky-Piscounov (sFKPP) equation \cite{KPP} well known in statistical physics. This analogy helps to understand most of the asymptotic properties of the evolution equation \cite{imm,it}.

Let us first consider the equation without noise (FKPP equation), equivalent to the BK equation in QCD \cite{Munier1}. This equation is known to admits travelling waves as asymptotic solutions \cite{vansaarloos}. In the framework of the BK equation, this means that at high rapidity, the solution takes the form of a travelling front at high $k$
\begin{equation}\label{eq:bkfront}
T(k,Y) \underset{Y\to\infty}{\approx} T\left(\frac{k}{Q_s(Y)}\right) \approx
\left(\frac{k}{Q_s}\right)^{-2\gamma_c},
\end{equation}
where
\begin{equation}\label{eq:bkqs}
Q_s^2(Y) \underset{Y\to\infty}{\approx} e^{v_c Y}
\end{equation}
is the saturation scale. Both the critical slope $\gamma_c\approx 0.6275$ and the critical velocity $v_c\approx 4.8836 \ab$ can be determined from the BFKL kernel $\chi(\gamma)$. In addition, the form of the traveling front looses traces of the initial condition\footnote{This is only true for steep enough initial conditions which, for the case of QCD, is ensured by colour transparency.}.

If we now switch on the noise term this picture gets modified and one has to consider the sFKPP equation. First of all, if one takes one single event of the Langevin equation, the asymptotic average velocity of the propagating front is smaller than $v_c$ \cite{brunet}
\begin{equation}\label{eq:speed}
v^* \underset{\alpha_s^2\kappa\to 0}{\to} v_c - \frac{\ab\pi^2\gamma_c\chi''(\gamma_c)}{2\log^2(\alpha_s^2\kappa)}.
\end{equation}
As indicated in the above equation, this analytic result is only valid for asymptotically small values of the noise strength $\alpha_s^2\kappa$. The shape of the front is also modified by the noise. It has even been shown \cite{vansaarloos} that, for the sFKPP equation, if we start with an initial condition which is always equal to 1 for $x<x_0$ and to 0 for $x>x_1$, then this compacity property is preserved by evolution. The extension of the tail is then proportional to $\log(\kappa)$. More precisely, the tail of the front behaves like
\[
A |\log(\alpha_s^2\kappa)| \sin\left[\frac{C\pi}{|\log(\alpha_s^2\kappa)|}\log\left(\frac{k^2}{Q_s^2(Y)}\right)\right] \exp\left[-\gamma_c\log\left(\frac{k^2}{Q_s^2(Y)}\right)\right].
\]

If we now consider a set of events, the effect of the noise will have a diffusive effect. This means that at rapidity $Y$, the events will show similar fronts distributed normally around $\bar Q_s(Y) = \avg{Q_s(Y)}$. The width of this dispersion is
\begin{equation}\label{eq:diffusion}
\Delta \log[Q_s^2(Y)] \approx \sqrt{D_{\text{diff}}\ab Y}
\end{equation}
typical of Gaussian white noises. It has been numerically found \cite{brunet} that, for asymptotically small $\alpha_s^2\kappa$ the front diffusion coefficient scales like
\[
D_{\text{diff}} \sim  \frac{1}{|\log^3(\alpha_s^2\kappa)|}.
\]

The point at this stage is that, although event by event geometric scaling is satisfied, if we average over multiple events, it is destroyed as we typically get an average amplitude behaving like a front traveling at a constant speed but spreading like $\sqrt{Y}$ \cite{imm}.

In the case of the equation \eqref{eq:langevin}, if one assume a front compact and saturating to a constant value, Iancu and Triantafyllopoulos have shown that the average value of the amplitude is
\begin{equation}\label{eq:frontit}
\langle T \rangle = \frac{N}{2} \erfc\left(\frac{z}{\sigma}\right)
+ \frac{N}{2}\exp\left(\frac{\gamma_c^2\sigma^2}{4}-\gamma_c z\right)\left[
2 - \erfc\left(\frac{z}{\sigma}-\frac{\gamma_c \sigma}{2} \right)
\right]
\end{equation}
where $z = \log(k^2/Q_s^2)$ is the position relative to the wavefront, $\sigma^2 \approx D_{\text{diff}}\ab Y$ is the diffusion of the events induced by fluctuations and $\erfc(x)$ is the complementary error function. The normalisation factor $N$ is given fixed to $\frac{1}{8}\chi''(1/2)\sqrt{1+8\chi(1/2)/\chi''(1/2)}\approx 6.98$ according to the transformation leading to the sFKPP equation. Finally, their model also predicts that, at asymptotic rapidities, the tail of the front satisfies
\begin{equation}\label{eq:avgeq}
\avg{T} \simeq \avg{T^2} \simeq \dots \simeq \avg{T^n} \simeq\dots
\end{equation}

It is important to note that our purpose is not only to test these predictions. The predictions \eqref{eq:speed} and \eqref{eq:diffusion}, obtained from the equivalence with the sFKPP equation, are only valid in the limit where the noise is very small {\em i.e.} in the limit of small and unphysical $\alpha_s$ (or, equivalently, small $\kappa$). In this work, we shall study the solutions of the QCD equation for physical values of the strong coupling. In addition, by analysing the complete QCD equation, without using the saddle point approximation, we shall also take into account the fact that, in the infrared region, the amplitude exhibits a logarithmic behaviour instead of saturating to a constant.

\section{Numerical solutions}\label{sec:numsol}

\subsection{Description of the method}

The aim of this paper is to solve numerically the evolution equation \eqref{eq:langevin} without simplifying it to the sFKPP equation and without using a model such as a inserting cut-off. To achieve this task we need two ingredients: an efficient way to deal with the linear-kernel convolution, and an efficient way to take into account the noise term. 

In order to observe the front propagation to large values of $k$ as energy increases, we shall work with the variable $L=\log(k^2/k_0^2)$ where $k_0$ is a fixed soft reference scale. The integration in \eqref{eq:langevin} is then performed by discretising space between $L_{\text{min}}$ and $L_{\text{max}}$. The compactness of the front in the presence of a noise term ensures that we can neglect the ultraviolet contribution coming from $T(L>L_{\text{max}})$. This is a crucial property which accelerates the convergence of the numerical algorithm w.r.t. the case of the BK equation. In the infrared region, one knows that the amplitude shows a logarithmic divergence {\em i.e.} $T(k) \approx \log(1/k)+\text{cst}$ for small $k$. Assuming that behaviour for $L<L_{\text{min}}$, one can compute analytically the contribution to the linear term coming from the infrared.

The treatment of the noise term is a little bit more tricky. The main problem is that, due to the presence of the square root in the pre-factor, one has to ensure that the noise does not produces negative values of $T$. The method we have used is the one proposed by Levine and Pechenik \cite{levine}. The technique is to consider the probability associated with the stochastic process. For the one under consideration, the probability that, in a rapidity interval $Y$, the amplitude goes from $T_0$ to $T$ satisfies the following equation
\[
\partial_Y P(Y,T;T_0) = \eta \partial_T^2 \left[ T P(Y,T;T_0)\right],
\]
with $\eta = \kappa \ab \as^2/(\pi\sqrt{\Delta L})$ ($\Delta L$ is the step for discretising the space dimension) and the initial condition
\[
P(0,T;T_0) = \delta(T-T_0).
\]
One can solve this differential equation and then impose that the amplitude remains positive. The final result is \cite{levine}
\begin{equation}\label{eq:proba}
P(Y,T;T_0) = \exp\left(-\frac{T_0}{\eta Y}\right) \delta(T)
 + \frac{1}{\eta Y} \sqrt{\frac{T_0}{T}} 
   \exp\left( -\frac{T+T_0}{\eta Y}\right) I_1\left(\frac{2TT_0}{\eta Y}\right).
\end{equation}
Once we know this probability, one can define the cumulative probability
\begin{equation}\label{eq:frq}
F_{T_0;\Delta Y}(T) = \int_{0^-}^T dT' \,P(\Delta Y,T';T_0),
\end{equation}
where the prescription on the inferior integration bound is such that the first term in \eqref{eq:proba} is included.

The method to apply the noise in then straightforward: if $u$ represent a uniform random variable on the interval $[0,1]$, $T \equiv F^{-1}_{T_0;\Delta Y}(u)$ is distributed according to the probability \eqref{eq:proba}. We can thus define an intermediate amplitude
\[
T_{\text{noise}}(k) = F^{-1}_{T(k);\Delta Y}[u(k)],
\]
where $u(k)$ is the uniformly-distributed random variable at position $k$. Once we have taken into account the fluctuation term we apply the remaining part of the evolution equation in order to obtain the evolved amplitude
\[
T(Y+\Delta Y, k) = T_{\text{noise}}(k) + \Delta Y \left[
\ab K\otimes T_{\text{noise}}(k) - \ab T_{\text{noise}}^2(k)\right]. 
\]

Finally, the integration \eqref{eq:frq} cannot be performed analytically. In order to simplify the numerical treatment of the noise term which already involves heavy computations, we have approximated the $T>0$ part of the probability \eqref{eq:proba} by a suitable Gaussian function leading to an error function for the cumulative frequency \eqref{eq:frq}.

In our numerical analysis, we shall perform discrete steps in rapidity according to \eqref{eq:ito}. It is thus important to notice that, due to the contribution at $T=0$, $F_{T_0;\Delta Y}(0) = \exp[-T_0/(\eta \Delta Y)]$ becomes close to 1 when $T_0 \ll \eta\Delta Y$. As a consequence, for sufficiently small values of $T$, the effect of the noise will be to set the amplitude to 0 with a probability almost equal to 1. This mechanism, having an effect close to a cut-off on $T$ as applied by Brunet and Derrida \cite{brunet} for the sFKPP equation, ensures the compacity of the front.

Before going on with the practical analysis, let us point out that, in order to reach very high values of the rapidity, we have, when necessary, worked in a {\em comoving frame}. This means that me have translated the value of $L_{\text{min}}$ and $L_{\text{max}}$ together with the propagating front. The value of $L_{\text{min}}$ has been adjusted in order to have $T(L_{\text{min}}) \ge 6$. The compactness of the front allows precise determination of the front in the ultraviolet region and the fact that the noise does not produce sizeable effects in the saturated region allows to handle the infrared part of the front.

\subsection{Simulation parameters}

We also have to discuss the discretisation of space and time. As mentioned earlier, the size of our steps in rapidity is related to the minimum value of the amplitude we want to describe. It is therefore important to set this to a small value. We have adopted a step size $\Delta Y=0.001$. The space discretisation is more complicated due to the presence of the Dirac distribution in $L=\log(k)$ in the noise dispersion. For a given discretisation of size $\Delta L$, the typical amplitude of the noise in one step of the evolution will be proportional to $\sqrt{\Delta Y/\Delta L}$. For the optimal stability in numerical simulations, it is thus safe to take a value of $\Delta L$ small enough but still larger than the time discretisation. We have obtained that $\Delta L=0.5$ was a good choice.

All our simulations have been performed with $\ab = 0.2$ which is typical for real-life situations and we have varied the strength $\alpha_s^2\kappa$ of the fluctuation term by simulating different values of $\kappa$. The studies at fixed coupling are relevant in this approach since there is no unique way to introduce running coupling effects at the leading order. In addition, by fixing $\alpha_s$, one could directly compare our results with the analytical predictions. For the purposes of numerical studies, we have worked with $\tilde\kappa = \frac{10 \pi}{N_c^2}\kappa$ instead of $\kappa$. 

The last thing that needs to be fixed is the initial condition. For all the following analysis, we have chosen a physically-motivated amplitude $T(Y=0)$ which already exhibits saturation in the infrared region:
\begin{equation}\label{eq:init}
T(Y=0,k) = \begin{cases}
\exp(-L) & \text{if }L>0,\\
1-\frac{L}{2} & \text{otherwise}.
\end{cases}
\end{equation}

\subsection{Results}

\begin{figure}
\subfigure[The dashed curves show 10 events generated with $\tilde\kappa=1$ at $Y=0,12.5,25,37.5$ and $50$ and the plain curve is their average.]{\includegraphics[scale=0.65]{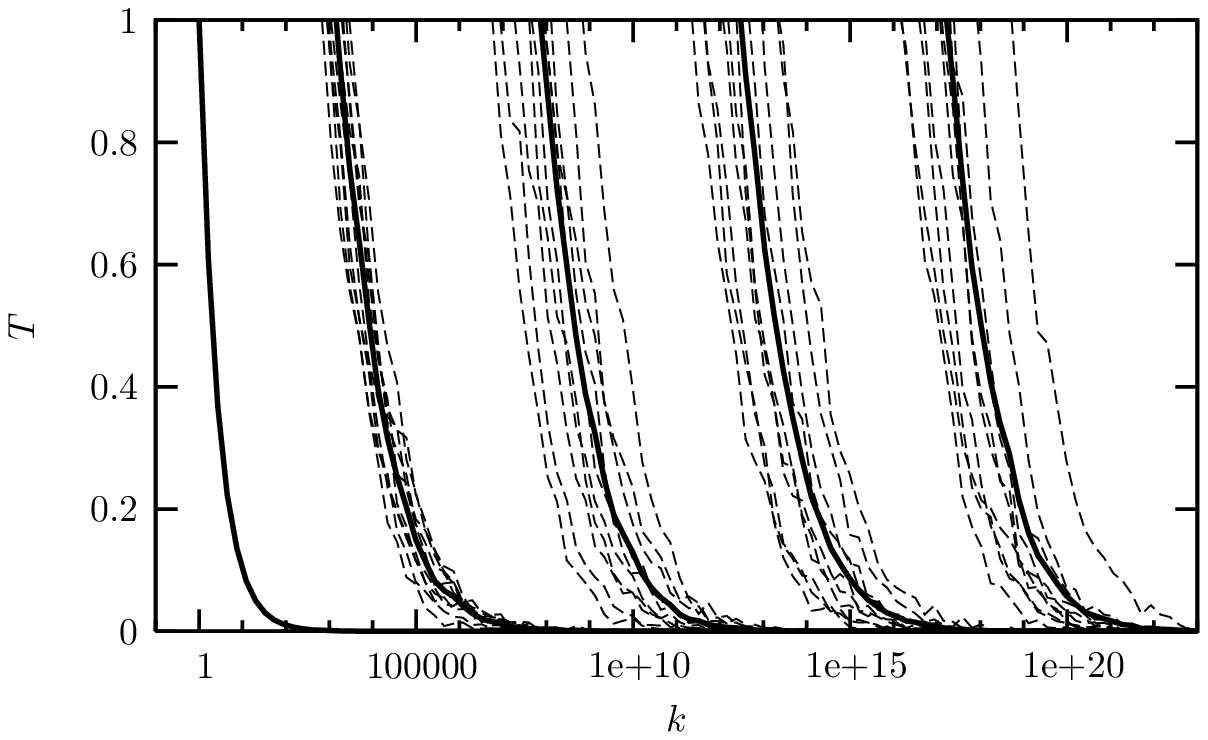}\label{fig:dispa}}
\subfigure[The points represent the square of the dispersion of the events as a function of rapidity together with a linear fit at high rapidities.]{\includegraphics[scale=0.65]{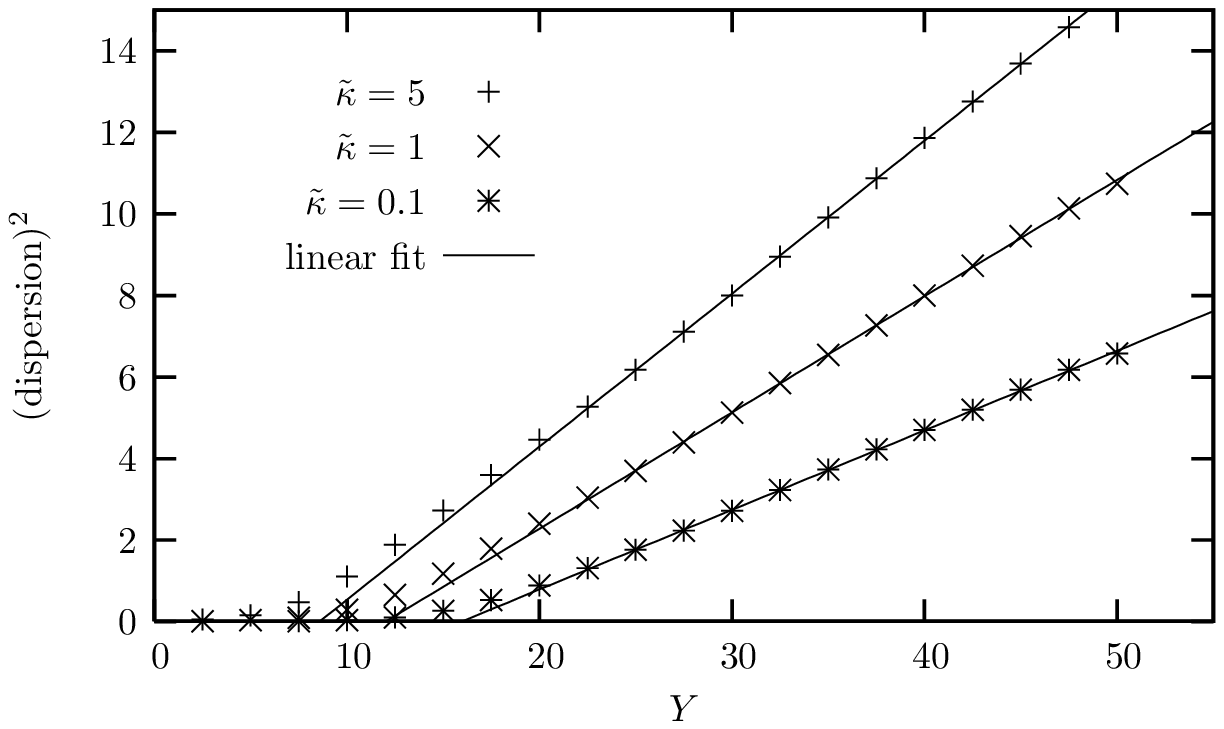}\label{fig:dispb}}
\caption{Dispersion of the events due to the noise effects as rapidity increases.}\label{fig:disp}
\end{figure}

\begin{figure}
\includegraphics{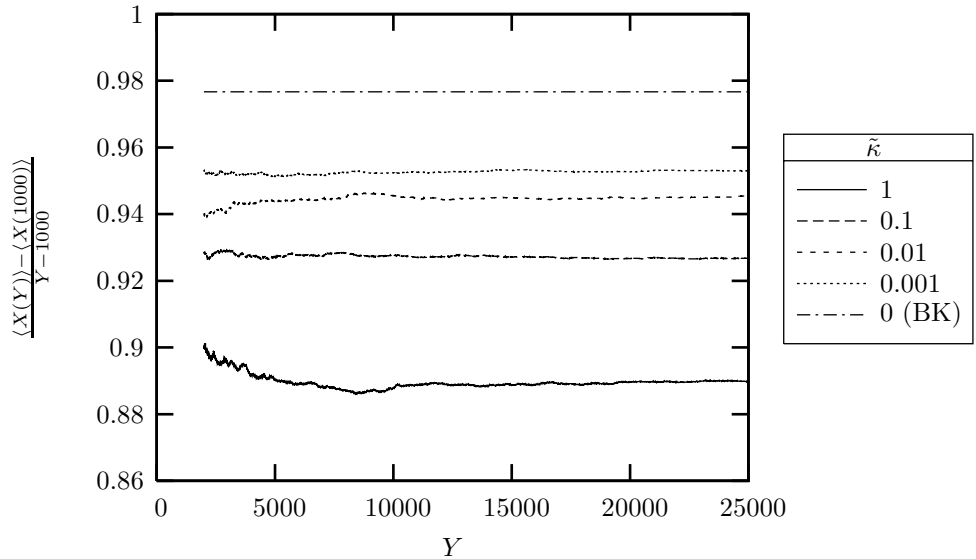}
\caption{Asymptotic speed of the traveling front. We observe a slow down w.r.t. the equation without fluctuations, more and more important as the strength of the noise is important.}\label{fig:speed}
\end{figure}

\begin{figure}
\includegraphics{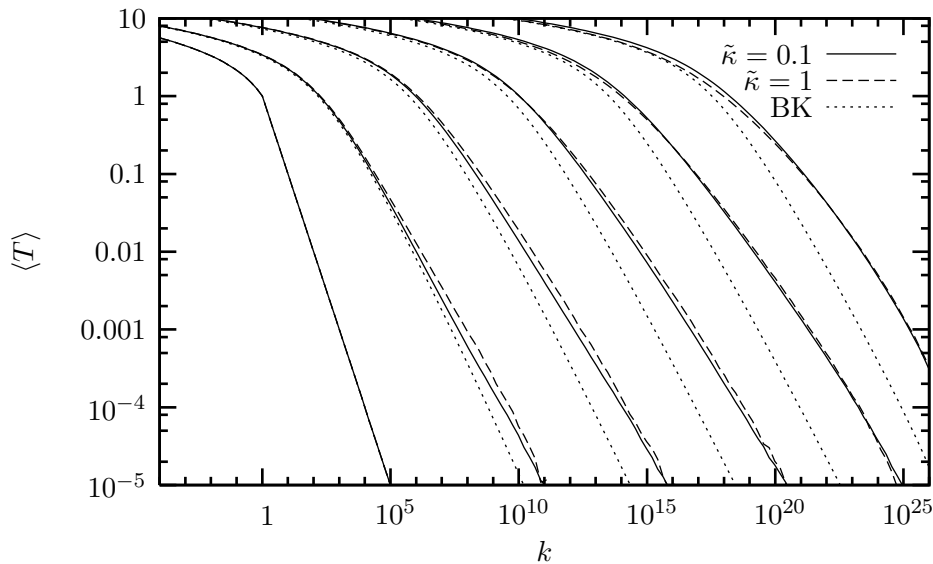}
\caption{Evolution with rapidity of the average value of the amplitude over 10000 events at $Y$ = 0,10,20,30,40 and 50.}
\label{fig:front_evol}
\end{figure}

\begin{figure}
\includegraphics{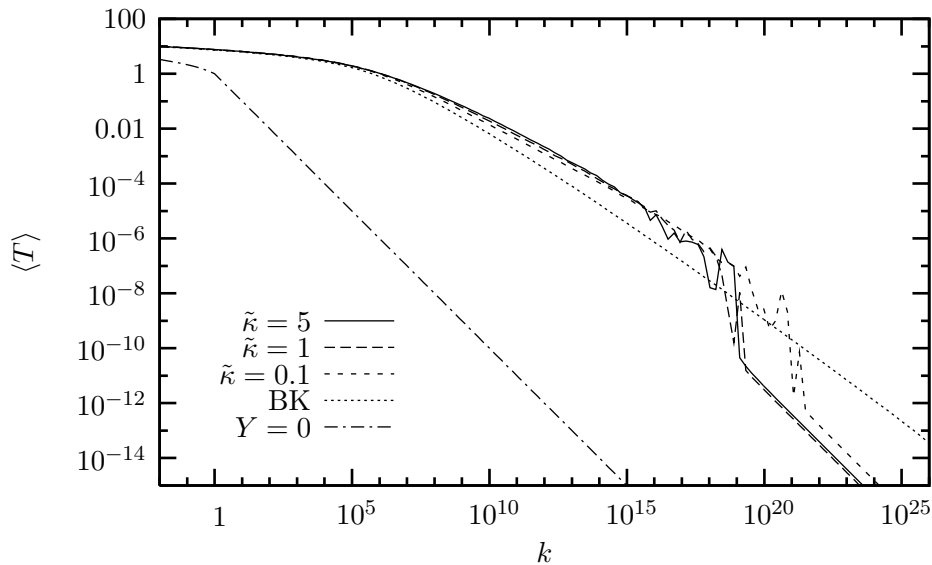}
\caption{Average value of the amplitude over 10000 events for $Y=20$ and different values of $\kappa$. The amplitude is shown for $Y=0,10,20,30,40$ and $50$.}
\label{fig:front_full}
\end{figure}

\begin{figure}
\includegraphics{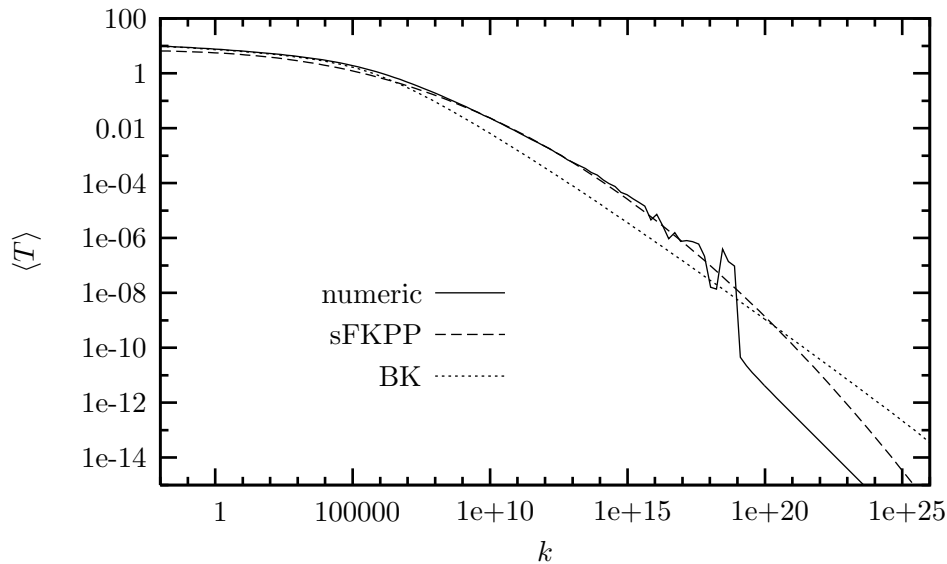}
\caption{Comparison of the numerical predictions for $\langle T \rangle$ at $Y=20$ with the amplitude obtained from BK evolution and with the sFKPP model \eqref{eq:frontit}. We have used $\tilde\kappa=1$.}
\label{fig:front_shape}
\end{figure}

\begin{figure}
\includegraphics{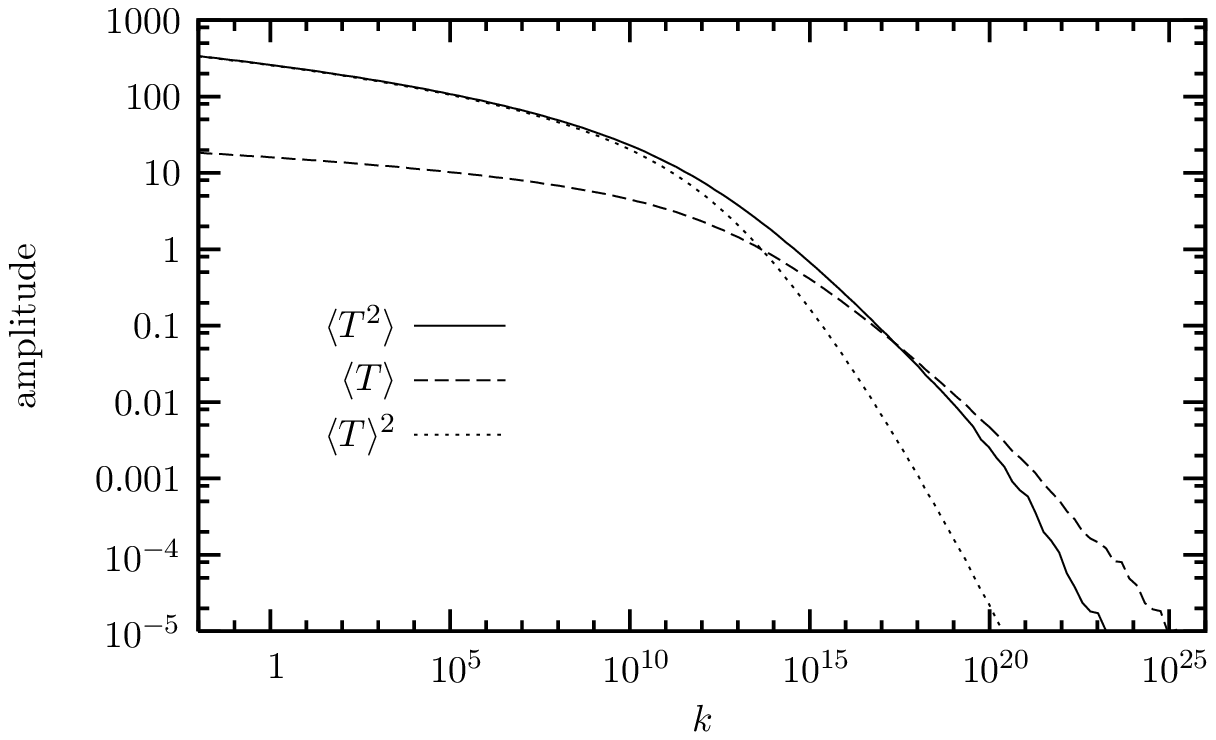}
\caption{Numerical results for the correlation $\langle T^2(k)\rangle$ at $Y=40$. We show $\langle T\rangle$ and $\langle T\rangle^2$ for comparison.}
\label{fig:front_t2}
\end{figure}

We have concentrated our studies on three major points. First, we have checked that, on an event-by-event basis, the dispersion of the front increases like $\sqrt{Y}$. Then, we have studied at very large rapidities the asymptotic velocity of the front. Finally, we have looked at the front averaged over a large number of events and we shall discuss its various properties.

We begin our analysis with the check that the position of the front is diffused like $\sqrt{Y}$. This diffusion is presented in figure \ref{fig:dispa} where we draw 10 events generated wit $\tilde\kappa=1$ together with their average for different values of the rapidity between 0 and 50. It is obvious that, as rapidity increases, these events become more and more spread. If we extract the position of each front by solving $T(L,Y) = 1/2$ w.r.t. $L$ for different values of $Y$, one can compute the statistical dispersion $\sigma$ of our set of events as a function of the rapidity. The result for $\sigma^2$ is drawn in figure \ref{fig:dispb} for a sets of 10000 events with different values of $\tilde\kappa$, together with a linear fit for $Y\ge 25$. This clearly proves that the dispersion of the front position goes like $\sqrt{Y}$ at large $Y$. Note that we have not gone to sufficiently low value of $\kappa$ to observe the asymptotic behaviour of the dispersion expected from \eqref{eq:diffusion}. Moreover, in the early stages of the evolution ($0\le Y\le 10$), we see that there is very few diffusion. In some sense, the diffusive process only starts to become important once the wavefront has reached its critical form. This point is very important since the dispersion of the events leads to violations of the geometric scaling. If, in the early stages of the evolution, we have very few diffusion, then geometric scaling will still be valid. This can thus explain why geometric scaling is indeed observed at HERA \cite{geometric} where $Y\lesssim 9$. Note however that this may of course depend on our choice of initial condition.

As a second test, we shall check that as $\tilde\kappa$ increases, the asymptotic velocity of the front decreases. Since the front velocity has rather large pre-asymptotic corrections, it is not sufficient, as for the determination of the diffusion, to consider a large number of events at $Y\approx 50$. The technique here is to use the comoving frame method to go to very large values of the rapidity. The speed can then be defined by
\[
v^*  = \lim_{Y\to\infty} \frac{\avg{L(Y)}-\avg{L(Y_0)}}{Y-Y_0}
\]
where $L(Y)$ is the position of the front at rapidity $Y$, determined by $T(L_s,Y)=1/2$, and $Y_0$ is a reference rapidity sufficiently high for the front to be stable. Practically, we have set $Y_0=1000$ and evolved the initial front up to $Y=25000$. Note that here, just a few events are needed in order to determine the velocity.
The results we obtained are presented in figure \ref{fig:speed} for different values of the noise strength $\tilde\kappa$. This figures shows clearly the fluctuations in speed and the slow down w.r.t. the BK equation where the fluctuations are not taken into account. This decrease of the speed becomes more and more important as the constant $\tilde\kappa$ grows. The logarithmic dependence on $\kappa$ cannot really be obtained from these analysis since we have not been to sufficiently small values of $\kappa$. This was however out of the scope of this paper since we simply want to check the decrease of the asymptotic speed when the noise becomes more and more important.
Note also that the set of 10000 events used in figure \ref{fig:dispb} for $\tilde\kappa=1$ gives a velocity of 0.912 at $Y=50$. Although this speed is smaller than the BK critical speed $v_c$, the asymptotic value \eqref{eq:speed} is not yet reached.

The main motivation of this paper was to study the shape of the traveling front  for physical strength of the noise term and its deviations from the results obtained in the absence of fluctuations. To do such an analysis, one need to generate a large number of events and take their average. 
Let us first study the evolution of the front with rapidity. We show in figure \ref{fig:front_evol} the amplitude\footnote{Note that the recoil observed in the last curve at the ultraviolet border of the domain is probably a non-physical numerical edge effect.} $\langle T \rangle$ obtained from 10000 events up to $Y=50$ and different values of the noise parameter $\tilde \kappa$. It is obvious on that figure that, as rapidity increases, the tail which is close to the BK result at small rapidities, starts to diverge from it due to diffusion in the events. The somehow surprising point is that, in the beginning of the evolution, the front including fluctuations travels faster than the BK front. This is not in contradiction with the fact that the asymptotic velocity needs to be smaller than the BK one, as we have seen previously that we need to go to much higher values of the rapidity in order to reach the asymptotic speed. The effect is really present in the early stages of the evolution and increases with $\tilde\kappa$ as we can see on the front for $Y=10$ and $Y=20$. Already at $Y=50$ one can observe a slow down of the front which is also more important for larger values of $\kappa$. It is also interesting to point out that analytical studies, only controlling the asymptotic behaviour (large $Y$ and weak noise), does not say anything about these early stages of the evolution.

To conclude our analysis, let us compare our simulations with the approximation introduced by Iancu and Triantafyllopoulos (see eqs. \eqref{eq:frontit} and \eqref{eq:avgeq}). In figure \ref{fig:front_shape}, we compare the numerical results for $\avg{T}$ with the front obtained from BK evolution and with equation \eqref{eq:frontit} where we have adjusted the parameters $Q_s$ and $\sigma$ in the range $10^9\le k\le 10^{18}$. We see a very good agreement between our results and the prediction \eqref{eq:frontit} for the tail of the front. The description in the saturated part of the front is not the same for the simple reason that the approximation \eqref{eq:frontit} assumes that the amplitude saturates to 1 instead of taking into account the logarithmic evolution which does not have any consequence on the high-$k$ behaviour. 

Finally, we have plotted in figure \ref{fig:front_t2} the average value of $T^2$ as a function of the momentum $k$ for $Y=40$ and $\tilde\kappa=1$, together with $\avg{T}$ and $\avg{T}^2$. In the small-$k$ region where the system is saturated, we clearly have $\avg{T^2}=\avg{T}^2$ which is coherent with the fact that we are in the dense regime and that the description is well reproduced also by the BK equation. In the dilute regime, we see that $\avg{T^2}$ becomes closer to $\avg{T}$ as predicted by \eqref{eq:avgeq}. The fact that the adequacy is not perfect simply comes from the reason that the asymptotic regime, strictly valid for $Y\to \infty$ is not completely reached. In that region, $T$ being smaller than 1 with a large probability, we have $T^2 < T$ and we expect $\avg{T^2}$ to reach $\avg{T}$ from below.

\section{Conclusions and discussion}\label{sec:ccl}

In this paper, we have studied numerically the solutions of the recently-proposed high-energy-QCD evolution equation including fluctuations. If we perform a coarse-graining in impact-parameter and go to momentum space, this hierarchy becomes equivalent to a Langevin equation. To analyse the effects due to the fluctuations, we have compared the results of the complete equation with the predictions from the BK evolution equation which only takes into account the BFKL and saturation contributions. Instead of solving the complicated infinite hierarchy of equations, we have simulated the equivalent Langevin problem which allows to study the QCD amplitudes on an event-by-event basis ant to obtain the average amplitude by generating a large number of events. Note that we have studied the complete QCD equation, with physically acceptable values of the parameters and without performing the approximations leading to the analogy with statistical physics.

Our analysis starts by showing that different events lead to a diffusion in position. More precisely, as rapidity increases, the positions of different events get spread with a dispersion proportional to $\sqrt(Y)$ and increasing with the importance of the noise term. In addition, it seems that this diffusion effect only becomes important once the wavefront is formed and hence we do not observe large dispersion in the early stages of the evolution. This may explain the fact that geometric scaling is indeed observed in experimental data as they lay in a region where dispersion appears to be small. This feature however may depend on our choice of initial condition \eqref{eq:init}. Indeed, to observe geometric scaling one has to be in a region of rapidity sufficiently large for the front to be formed and not too large so that fluctuations effects are not yet important. At asymptotically weak noises, the time of formation of the front is of order $\beta^{-1}\gamma_c^{-2}\log(\kappa \alpha_s^2)$ with $\beta\approx 48.2$. This is faster than the equivalent time for the BK equation and one might expect that geometric scaling sets in early. However, this is only known for asymptotically weak noises while the effect of fluctuations is only known at asymptotically high rapidities. Therefore, very few things can be predicted concerning these competitive effects and complementary studies are required.

We have then studied the asymptotic speed of the traveling front {\em i.e.} the rapidity evolution of the saturation scale. This requires to go to very large values of the rapidity in order not to be sensitive to the sub-asymptotic corrections and to the diffusion due to the noise which introduce corrections proportional to $Y^{-1/2}$. As predicted by the sFKPP equation, the asymptotic saturation exponent is significantly smaller than the one obtained from the BK equation. Note however that since we have not studied asymptotically small values of the noise strength parameter $\alpha_s^2\kappa$ which is out of the scope of this paper, we have not observed the $\log^{-2}(\alpha_s^2\kappa)$ behaviour of the corrections. It appears however that the slow down of the front predicted from statistical physics still holds in the QCD case.

Finally, we have studied the interesting problem of the shape of the travelling front in the presence of fluctuations. This is performed by considering a large number of events and comparing the results, with different values of the noise parameter $\kappa$, with the front obtained from BK evolution. We have observed at large rapidities the violations of geometric scaling in the leading part of the tail and, at very large momentum $k$, a strong damping in agreement with the compacity property of the front in such type of stochastic evolution equation. We have observed that, in the early stage of the evolution the front goes faster than the front obtained from BK evolution. At higher rapidities, we already observe that the speed of the average front goes slower than the front without noise term. This speed also turns out to be smaller for larger contributions from fluctuations. Moreover, we have checked that the front predicted by the stochastic QCD evolution shows a behaviour consistent with the form predicted by Iancu and Triantafyllopoulos. At high rapidities, we have checked that, according to the theoretical predictions, the average values $\avg{T^2}$ behaves like $\avg{T}^2$ in the saturated region and converges to $\avg{T}$ in the tail of the traveling front.

In the future, it would be interesting to see if these numerical studies can help to distinguish some effects due to fluctuations in the experimental data. To perform these phenomenological studies, one has to consider the effects of running coupling. One may notice that the compacity property of the traveling front gives an efficient way to introduce running coupling in the evolution equation. Indeed, the dynamics of the evolution is mainly located in the region of the front and thus one should use $Q_s^2$ as the scale for the running coupling. As for the case of the BK equation, one expects a front evolving like $\sqrt{Y}$ instead of $Y$. The diffusion of the events as a function of rapidity is unclear and requires more precise analyses.

Then, one may also introduce the impact-parameter dependence which is known to lead to a stochastic equation with a multi-variable noise, poorly known in theoretical grounds. However, if one consider the BK equation, the introduction of the impact-parameter dependence \cite{mars} does not modify the asymptotic properties observed in the $b$-independent case. Thus we do not expect that the full hierarchy will lead to huge qualitative modifications of the present picture.

At the level of purely theoretical studies, our results suggest to analyse the link with the DLA regime which appears to be closely related with the compacity properties of the front.

\begin{acknowledgments}
  I would like to thank E. Iancu and D. Triantafyllopoulos for very stimulating discussions and for encouraging this work. I am also very grateful to E. Iancu and R. Peschanski for carefully proofreading the manuscript. The author is funded by the National Funds for Scientific Research (Belgium).
\end{acknowledgments}

\begin{appendix}\label{app:equiv}
\section{Equivalence between the Langevin equation and the hierarchy}

Instead of considering the full QCD case, we shall show how a Langevin equation translates into an infinite hierarchy of equation of the same form as \eqref{eq:hier} in a simplified model. Since the introduction of space variable with sophisticated linear kernel simply introduces complications with the notations, let us consider the 0-dimensional problem. 

Our starting point is the Langevin equation
\[
\partial_t u(t) = u(t) - u^2(t) + \sqrt{2\alpha u(t)} \eta(t),
\]
with the noise $\eta(t)$ satisfying $\langle\eta(t)\rangle=0$ and $\langle\eta(t)\eta(t')\rangle=\delta(t-t')$. Since a Langevin process is not differentiable, one has to understand this equation as the continuum limit with an Ito prescription. If we discretise the time in intervals of size $\varepsilon$, it gives
\[
\frac{u_{j+1}-u_j}{\varepsilon} = u_j - u_j^2 + \sqrt{2\alpha u_j} \eta_j,
\]
where the noise satisfying
\[
\langle\eta_j\rangle=0\quad\text{and}\qquad \langle\eta_j\eta_{j'}\rangle=\frac{1}{\varepsilon}\delta_{jj'}.
\]
Let us now consider the time evolution of the average value $\langle F(u)\rangle$ of a function $F(u)$. Using again the Ito prescription and expanding $F(u_{j+1})$ to second order around $u_j$, it is straightforward to get
\[
\partial_t \langle F(u) \rangle = \langle uF'(u) \rangle - \langle u^2F'(u) \rangle + \alpha \langle u F''(u) \rangle.
\]
Taking $F(u) = u^n$ and defining $u^{(k)} = \langle u^k\rangle$ we obtain the infinite hierarchy of equations
\[
\partial_t u^{(k)} = ku^{(k)} - ku^{(k+1)} + \alpha k(k-1) u^{(k-1)},
\]
which reproduces a linear, a merging and a splitting term.

\end{appendix}

\end{document}